\documentclass{JHEP3} 
\usepackage{amsmath}

\title{String breaking in QCD: dual superconductor {\it vs.} stochastic vacuum model}

\author{D. Antonov
        \thanks{Permanent address: 
        ITEP, B. Cheremushkinskaya 25, RU-117 218 Moscow, Russia.}\\ 
        Institute of Physics, Humboldt University of Berlin,\\
        Newtonstr. 15, 12489 Berlin, Germany \\       
        E-mail: \email{antonov@physik.hu-berlin.de}}

\author{A. Di Giacomo\\
Dipartimento di Fisica ``E. Fermi'' dell'Universit\'a di Pisa,\\
INFN-Sezione di Pisa,\\
Largo Pontecorvo, 3, 56127 Pisa, Italy \\
E-mail: \email{digiaco@df.unipi.it}}

\abstract{Effects of dispersion of the chromoelectric field of the flux tube on the 
string-breaking distance are studied. The leading-order correction is shown to slightly diminish the result
following from the Schwinger formula. Instead, accounting for corrections of all orders might
result, at certain values of the Landau-Ginzburg parameter, in an increase of the 
string-breaking distance up to one order of magnitude. 
An alternative formula for this distance is obtained when produced pairs are treated as 
holes in a confining pellicle, which spans over the contour of an external quark-antiquark pair. Generalizations of the 
obtained results to the cases of small temperatures, as well as temperatures close to the critical one are also discussed.}


\preprint{HU-EP-05/01 \\
IFUP-TH 2005/01} 

\dedicated{}

\begin{document}

\section{Introduction. The model and some relations between its parameters}
String breaking at zero and finite temperatures is known as one of the important open problems in QCD (see e.g. refs.~\cite{a,b,c,d,e,1,f}
for recent developments).
The aim of this paper is to proceed further with the calculation of the string-breaking distance, along with the 
lines of ref.~\cite{1} (Sections~2-6), as well as using some alternative model (Section~7). In ref.~\cite{1},
dual Abrikosov-Nielsen-Olesen strings have been used to model flux tubes of the chromoelectric field in real QCD. 
Such strings~\cite{15} 
are solutions to the classical equations of motion in the 4d dual Abelian Higgs model, whose Euclidean Lagrangian reads
${\cal L}=\frac14F_{\mu\nu}^2+|D_\mu\varphi|^2+\frac{\lambda}{2}(|\varphi|^2-v^2)^2$. Here 
$F_{\mu\nu}=\partial_\mu B_\nu-\partial_\nu B_\mu$, $D_\mu\varphi=(\partial_\mu-ig_mB_\mu)\varphi$,
$B_\mu$ is the dual gauge field, $\varphi$ is the complex-valued dual-Higgs field, and $g_m$ is the magnetic 
coupling constant, related to the electric one, $g$, as $g_m=2\pi/g$. The masses of the dual vector boson and 
the dual Higgs boson are $m_V=\sqrt{2}g_m v$ and $m_H=\sqrt{2\lambda}v$, respectively. 

We will consider this model in the 
London limit~\footnote{The ratio $\frac{m_H}{m_V}$ is called Landau-Ginzburg parameter.}, 
$L\equiv\ln\frac{m_H}{m_V}\gg 1$, where all the 
results can be obtained analytically. The Bogomol'nyi case, $L=0$, will be considered elsewhere. 
The electric field of a straight-line dual Abrikosov-Nielsen-Olesen string reads $E(r)=\frac{m_V^2}{g_m}K_0(m_Vr)$, 
where $r=|{\bf x}_\perp|$ with ${\bf x}_\perp=(x_1, x_2)$ denoting 
the direction transverse to the string, and from now on $K_\nu$'s stand for 
MacDonald functions.
The field averaged over the string cross section, 
$\left<E\right>=\frac{1}{S}\int d^2r E(r)$ obeys the relation $g\left<E\right>=4\sigma/L$. Here 
$S=\pi m_V^{-2}$ is the area of the cross section of the string,
and the string tension has the form $\sigma=2\pi v^2L$. In order to establish a 
correspondence to QCD, we will need to express all the results in terms of $\sigma$ and $g$. Two formulae useful for this 
are $m_V^2=\frac{4\pi\sigma}{g^2L}$ and $S=\frac{g^2L}{4\sigma}$.

\section{The string-breaking distance with the neglection of dispersion of $E(r)$}

The rate of the pair production reads $w=\frac{2}{S}{\,}{\rm Im}{\,}\Gamma[A_i]$, 
where the one-loop effective action of a scalar quark
has the form:

$$
\Gamma[A_i]=N_cN_f\int\limits_{0}^{\infty}\frac{dT}{T}{\rm e}^{-m^2T}\times$$

\begin{equation}
\label{1}
\times\int {\cal D}{\bf x}_\perp {\cal D}{\bf x}_\parallel
\exp\left[-\int\limits_{0}^{T}d\tau\left(\frac14\dot{\bf x}_\perp^2+\frac14\dot{\bf x}_\parallel^2-\frac{ig}{2}
E({\bf x}_\perp(\tau))
\epsilon_{ij}\dot x_i x_j\right)\right].
\end{equation}
Here, ${\bf x}_\parallel=(x_3, x_4)$, $i,j=3,4$, 
and the field of the string is $A_i=-\frac12\epsilon_{ij}x_jE({\bf x}_\perp)$. Note that, 
in this paper, we will study the case of scalar quarks only (except of some comment at the end of the next Section), since  
the longitudinal part of the spin factor, $\int{\cal D}\psi_i\exp\left[-\int\limits_{0}^{T}d\tau\left(\frac12\psi_i
\dot\psi_i-ig\psi_i\psi_j F_{ij}(\tau)\right)\right]$, 
where $F_{34}(\tau)=-F_{43}(\tau)=E({\bf x}_\perp(\tau))$, cannot be calculated exactly
as long as $F_{ij}$ is $\tau$-dependent. To evaluate the path integral~(\ref{1}), we will use the requirement that the 
mass of the produced pair, $m$, should be much larger than $m_V$, i.e. $m\gg\frac2g\sqrt{\frac{\pi\sigma}{L}}$,
in order that the field of a string can be considered as a constant one. The ``largeness'' of $m$ means that characteristic
proper times are ``small'', $T<\frac{1}{m^2}$, that enables us to evaluate the path integral semiclassically and to 
compute further the leading quantum correction using the 
Feynman variational method~\cite{2}. Furthermore, we naturally assume that not only the Compton wavelength of a
produced pair, $m^{-1}$, is much smaller than the range of the field localization, $m_V^{-1}$, but also that the 
characteristic pair trajectories are small compared to $m_V^{-1}$. Since, in the Euclidean space-time, pair trajectories
are circles of the radius $R_p=\frac{m}{g\left<E\right>}$~\footnote{This can be seen either by solving the respective
Euler-Lagrange equation~\cite{3}, or simply by noticing that $\exp\left(-\frac{\pi m^2}{g\left<E\right>}\right)$ in the 
Schwinger formula should be ${\rm e}^{-\Phi}$, where $\Phi$ is the flux of the electric field through the contour 
of a pair, $\Phi=g\left<E\right>\cdot \pi R_p^2$. Therefore, a pair is identified with a quark which moves along a 
circle of the radius $R_p$.}, the condition of smallness of the pair trajectory, $R_p\ll m_V^{-1}$,
yields $m\ll 2g\sqrt{\frac{\sigma}{\pi L}}$. Both conditions, 

\begin{equation}
\label{ineq}
\frac2g\sqrt{\frac{\pi\sigma}{L}}\ll m\ll 2g\sqrt{\frac{\sigma}{\pi L}},
\end{equation}
are compatible to each other at $g\gg 1$. 

Then, due to the smallness of 
a pair trajectory, $E({\bf x}_\perp(\tau))$ can be replaced by its value averaged along the trajectory:

\begin{equation}
\label{smallT}
\int\limits_{0}^{T}d\tau E({\bf x}_\perp(\tau))\dot x_i x_j\simeq -\Sigma_{ij}
\cdot\frac{1}{T}\int\limits_{0}^{T}d\tau E({\bf x}_\perp(\tau)),
\end{equation} 
where $\Sigma_{ij}\equiv\int\limits_{0}^{T}d\tau x_i \dot x_j$ is the $(i,j)$-th component of the so-called tensor area 
of the trajectory. In the leading small-$T$ approximation, we obtain the classical result for $\int {\cal D}{\bf x}_\perp=
\int d^2x_\perp(0)\int\limits_{{\bf x}_\perp(0)={\bf x}_\perp(T)}^{}{\cal D}{\bf x}_\perp(\tau)$
in eq.~(\ref{1}): 

\begin{equation}
\label{class}
\frac{1}{4\pi T}\int d^2x_\perp\exp\left[-\frac{ig}{2}E({\bf x}_\perp)\epsilon_{ij}\Sigma_{ij}\right].
\end{equation}
Thus,

\begin{equation}
\label{eff}
\Gamma[A_i]\simeq\frac{N_cN_f}{4\pi}\int\limits_{0}^{\infty}\frac{dT}{T^2}{\rm e}^{-m^2T}\int {\cal D}{\bf x}_\parallel
\exp\left(-\frac14\int\limits_{0}^{T}d\tau\dot{\bf x}_\parallel^2\right)\int d^2x_\perp\exp\left[-\frac{ig}{2}E({\bf x}_\perp)
\epsilon_{ij}\Sigma_{ij}\right].
\end{equation}
Neglecting for this Section the dispersion of the field $E({\bf x}_\perp)$, we have

\begin{equation}
\label{0}
\int d^2x_\perp\exp\left[-\frac{ig}{2}E({\bf x}_\perp)
\epsilon_{ij}\Sigma_{ij}\right]\equiv S\left<\exp\left[-\frac{ig}{2}E({\bf x}_\perp)
\epsilon_{ij}\Sigma_{ij}\right]\right>\simeq S\exp\left[-\frac{ig}{2}\left<E\right>
\epsilon_{ij}\Sigma_{ij}\right].
\end{equation}
In this approximation, we therefore 
arrive at the Euler-Heisenberg Lagrangian in the constant field $\bar A_i\equiv-\frac12\epsilon_{ij}x_j
\left<E\right>$,

\begin{equation}
\label{effact}
\Gamma[A_i]\simeq S\frac{N_cN_f}{(4\pi)^2}\int\limits_{0}^{\infty}\frac{dT}{T^2}{\rm e}^{-m^2T}
\frac{g\left<E\right>}{\sin(g\left<E\right>T)},
\end{equation}
and recover for $w$ the 4d Schwinger result~\footnote{Up to the factor $N_cN_f$ absent
in the electromagnetic case.} in the bosonic case:

\begin{equation}
\label{w}
w=N_cN_f\frac{(g\left<E\right>)^2}{(2\pi)^3}\sum\limits_{k=1}^{\infty}\frac{(-1)^{k+1}}{k^2}
\exp\left(-\frac{\pi k m^2}{g\left<E\right>}\right).
\end{equation}
We can further express the inequality $T<\frac{1}{m^2}$ in terms of the parameters of our model.
Namely, since at least the first pole in the imaginary part of the Euler-Heisenberg Lagrangian should give its 
contribution, $T$ may not be arbitrarily small, but the following inequality should rather hold: 
$T>\frac{\pi}{g\left<E\right>}=\frac{\pi L}{4\sigma}$. The condition $\frac{1}{m^2}>T$ then yields
$m<2\sqrt{\frac{\sigma}{\pi L}}$, that also coincides with 
the condition for $w$ not to be exponentially small. This new constraint 
is stronger than the above-obtained one, expressed by the right inequality of~(\ref{ineq}),
since $g\gg 1$ is now absent.
The new constraint can be viewed as an upper boundary on $L$: 

\begin{equation}
\label{L}
L<\frac{4}{\pi}\frac{\sigma}{m^2}\simeq 1.27\frac{\sigma}{m^2}.
\end{equation} 
Setting $\sigma=(440{\,}{\rm MeV})^2$ and a typical hadronic mass $m=200{\,}{\rm MeV}$, 
we get an estimate $L<6.2$, that still leaves a window for $L\gg 1$.

Approximating the whole sum~(\ref{w}) by its first term (equal to the density of produced pairs), we have

\begin{equation}
\label{ww}
w\simeq\frac{2N_cN_f}{\pi^3}\left(\frac{\sigma}{L}\right)^2\exp\left(-\frac{\pi
m^2L}{4\sigma}\right).
\end{equation}
The respective string-breaking distance~\cite{1} $\bar r=\frac{1}{\sqrt{2Sw}}$ has the form

\begin{equation}
\label{rr}
\bar r=\frac{\pi^{3/2}\sqrt{L}}{g\sqrt{N_cN_f\sigma}}\exp\left(\frac{\pi L}{8}\frac{m^2}{\sigma}\right),
\end{equation}
where, due to inequality~(\ref{L}),
$\exp\left(\frac{\pi L}{8}\frac{m^2}{\sigma}\right)<\sqrt{{\rm e}}\simeq 1.65$.

\section{Finite-temperature generalizations}
Setting for the mass of a produced pair the $\pi$-meson mass, 
we can further extend the analysis of the previous Section to the case where the temperature is close to the critical one.
This can be done by using the formulae~\cite{1, 4} $m^2=m_\pi^2t^{1.44}$,
$\sigma=\sigma_0t^{0.33}$, where $t\equiv 1-\frac{T}{T_c}$ is the reduced temperature~\footnote{The fact that 
the temperature and the proper time are denoted by the same letter ``$T$'' should not lead to reader's confusions.}, and 
$\sigma_0\simeq (440{\,}{\rm MeV})^2$.
We obtain:

\begin{equation}
\label{tto0}
w\to\frac{2N_cN_f}{\pi^3}\left(\frac{\sigma}{L}\right)^2={\cal O}\left(t^{0.66}\right),
\end{equation}
that establishes the law by which $\bar r$ grows at $t\to 0$.
In the same limit $t\to 0$, the condition $mr_\perp\gg 1$ with 
the temperature-dependent $r_\perp$, $r_\perp=\frac{g}{2}\sqrt{\frac{L}{\pi\sigma}}$, leads to the following 
boundary on $t$ from below: $\frac1L\left(\frac{1}{m_\pi}\sqrt{\frac{\sigma_0}{\alpha_s}}\right)^{1.79}\ll t$.
On the other hand,  
the condition of smallness of the 
pair trajectory, expressed by the right inequality of~(\ref{ineq}), sets a boundary on $t$ from above: 
$t\ll\frac{1}{L^{0.91}}\left(\frac{2g}{m_\pi}\sqrt{\frac{\sigma_0}{\pi}}\right)^{1.82}$. 
These two conditions imposed on $t$ are apparently compatible to each other at sufficiently large $L$ and/or $g$.
Note also that the condition~(\ref{L}) becomes
softer as one approaches the critical point, since $\frac{\sigma}{m^2}=\frac{\sigma_0}{m_\pi^2}t^{-1.11}\to\infty$.

Let us now evaluate the string-breaking 
distance at relatively low temperatures, namely at $T={\cal O}(f_\pi)$, where the following 
formula holds~\cite{GL}:

\begin{equation}
\label{mT}
m_\pi^2(T)\simeq m_\pi^2\left(1+\frac{T^2}{24f_\pi^2}\right).
\end{equation} 
We can further approximate
$m_V(T)$ by $m_V(0)$, since $m_V^{-1}$ is the vacuum correlation length, whose QCD-analogue at such temperatures
can be considered as temperature-independent \cite{correl} \footnote{In QCD, the (magnetic) vacuum 
correlation length becomes definitely temperature-dependent at temperatures larger than the temperature 
of dimensional reduction (that is of the order of $2T_c$), where it is proportional to $\frac{1}{g^2(T)T}$.}.

Next, we need to know the temperature dependence of $g\left<E\right>$. Since we are exploring the region of 
temperatures smaller than the temperature of dimensional
reduction, the sum over Matsubara frequencies appears:

$$g\left<E\right>(T)=\frac{(gm_V)^2}{2\pi^2}\int d^2z\sum\limits_{n=-\infty}^{+\infty}
K_0\left(\sqrt{z^2+(m_V\beta n)^2}\right),$$
where $\beta\equiv 1/T$. To carry out the integral, it is convenient to transform 
the sum as follows:

$$
\sum\limits_{n=-\infty}^{+\infty}K_0\left(\sqrt{z^2+(m_V\beta n)^2}\right)=\frac12\sum\limits_{n=-\infty}^{+\infty}
\int\limits_{0}^{\infty}\frac{dt}{t}\exp\left[-\frac{1}{4t}-t\left(z^2+(m_V\beta n)^2\right)\right]=$$

\begin{equation}
\label{transform}
=\frac{T\sqrt{\pi}}{2m_V}\sum\limits_{n=-\infty}^{+\infty}\int\limits_{0}^{\infty}\frac{dt}{t^{3/2}}
\exp\left[-z^2t-\frac{1+(\omega_n/m_V)^2}{4t}\right]=
\frac{\pi T}{m_V}\sum\limits_{n=-\infty}^{+\infty}
\frac{{\rm e}^{-|z|\sqrt{1+(\omega_n/m_V)^2}}}{\sqrt{1+(\omega_n/m_V)^2}},
\end{equation}
where $\omega_n\equiv2\pi Tn$. 
The integration over $d^2z$ then immediately yields 

\begin{equation}
\label{gET}
g\left<E\right>(T)=g^2m_VT\sum\limits_{n=-\infty}^{+\infty}
\frac{1}{\left[1+(\omega_n/m_V)^2\right]^{3/2}}.
\end{equation}
To study the zero-temperature limit of this expression, one should perform the inverse transformation of the sum:

$$
\sum\limits_{n=-\infty}^{+\infty}
\frac{1}{\left[1+(\omega_n/m_V)^2\right]^{3/2}}=\frac{2}{\sqrt{\pi}}
\int\limits_{0}^{\infty}dt\sqrt{t}
{\rm e}^{-t}\sum\limits_{n=-\infty}^{+\infty}{\rm e}^{-t(\omega_n/m_V)^2}=$$ 

$$=\frac{m_V\beta}{\pi}\sum\limits_{n=-\infty}^{+\infty}
\int\limits_{0}^{\infty}dt\exp\left[-t-\frac{(m_V\beta n)^2}{4t}\right]=
\frac{(m_V\beta)^2}{\pi}\sum\limits_{n=-\infty}^{+\infty}|n|K_1(m_V\beta |n|).$$
At small $T$'s of interest, the sum here is apparently dominated by the zeroth mode.
Moreover, expanding the function $K_1$ in power series, one can check that no finite corrections exist 
to the r.h.s. of the formula $|n|K_1(m_V\beta |n|)\stackrel{|n|\to 0}{\longrightarrow}T/m_V$. Therefore,
the leading small-$T$ (physically, at $T\ll f_\pi$) correction to $g\left<E\right>(T)$
stems from the terms in the sum with $|n|=1$, $2K_1(m_V\beta)$. This correction is therefore 
exponentially small, as confirmed by the following final expression:

\begin{equation}
\label{GE}
g\left<E\right>(T)\simeq\frac{4\sigma_0}{L}\left[1+2\sqrt{\frac{\beta}{g}}\left(\frac{\pi^3\sigma_0}{L}\right)^{1/4}
{\rm e}^{-\frac2g\sqrt{\frac{\pi\sigma_0}{L}}\beta}\right],
\end{equation}
where $\sigma_0\equiv2\pi v^2L$. As for the dependence $\sigma(T)$, it is derived in Appendix~B.

Equations~(\ref{mT}) and (\ref{GE}), being substituted into the formula
$w\simeq N_cN_f\frac{(g\left<E\right>)^2}{(2\pi)^3}{\rm e}^{-\frac{\pi m^2}{g\left<E\right>}}$,
determine the temperature dependence of $w$. Apparently, the correction produced by eq.~(\ref{GE})
is exponentially small with respect to the temperature dependence appearing by means of eq.~(\ref{mT}). Therefore, 
$w$ decreases at $T$ increasing from zero to the temperatures of the order of $f_\pi$.
The corresponding increase of $\bar r$ parallels the same phenomenon we have found at $T\to T_c$.

It is finally worth making the following comment. In the case where the dispersion of $E(r)$ is 
neglected (that we are discussing here), the spin 
factor can be evaluated, and it is natural to address the issue of how strongly the antiperiodic boundary conditions 
for spin-$\frac12$ quarks can 
affect the obtained result. For such quarks, one should substitute in eq.~(\ref{effact}) $\frac{1}{\sin(g\left<E\right>T)}\to
-\frac{2}{\tan(g\left<E\right>T)}$, that yields, instead of eq.~(\ref{w}),

\begin{equation}
\label{Wferm}
w=N_cN_f\frac{(g\left<E\right>)^2}{4\pi^3}\sum\limits_{k=1}^{\infty}\frac{1}{k^2}
\exp\left(-\frac{\pi k m^2}{g\left<E\right>}\right).
\end{equation}
It has been shown in ref.~\cite{bras} that 
antiperiodic boundary conditions for fermions can be taken into account upon the multiplication 
of the zero-temperature one-loop effective action by the factor
$\left[1+2\sum\limits_{n=1}^{\infty}(-1)^n{\rm e}^{-\frac{\beta^2n^2}{4T}}\right]$, where (only for this formula in this Section)
$T$ stands for the 
proper time. In course of taking ${\rm Im}{\,}\Gamma[A_i]$, this factor transforms to another one, by which a $k$-th
term of the series in eq.~(\ref{Wferm}) should be multiplied: 
$\left[1+2\sum\limits_{n=1}^{\infty}(-1)^n\exp\left(-\frac{g\left<E\right>\beta^2n^2}{4\pi k}\right)\right]$.
For relevant $k$'s, which are $k<\frac{g\left<E\right>}{\pi m^2}$, we obtain the following 
relevant $n$'s: $n<2T\sqrt{\frac{\pi k}{g\left<E\right>}}=\frac{2T}{m}$, that, at $T<f_\pi$, are 
smaller than 2. Therefore, the factor produced by the antiperiodic boundary conditions 
for spin-$\frac12$ quarks reduces to
$\left[1-2\exp\left(-\frac{g\left<E\right>\beta^2}{4\pi k}\right)\right]$. Since, for the above-mentioned relevant $k$'s, 
$\frac{g\left<E\right>\beta^2}{4\pi k}>\left(\frac{m\beta}{2}\right)^2$,
at $T<\frac{m_\pi}{\sqrt{2}}$ the obtained correction falls off as the tail of the 
Gaussian distribution~\footnote{The inequality $T<\frac{m_\pi}{\sqrt{2}}$ stems from the fact that the 
width of the Gaussian distribution ${\rm e}^{-m^2\beta^2/4}$ is $\frac{\sqrt{2}}{m}$.}.

\section{Significance of corrections due to the dispersion of $E(r)$}
Let us now take into account the second cumulant in eq.~(\ref{0}) [i.e. the dispersion of $E(r)$] 
and also address the issue of convergence of the cumulant expansion. We have

$$\left<\exp\left[-\frac{ig}{2}E({\bf x}_\perp)
\epsilon_{ij}\Sigma_{ij}\right]\right>\simeq \exp\left[
-\frac{ig}{2}\left<E\right>
\epsilon_{ij}\Sigma_{ij}-\frac{g^2}{8}(\epsilon_{ij}\Sigma_{ij})^2\left(\left<E^2\right>-\left<E\right>^2\right)\right]=$$

\begin{equation}
\label{A}
=\frac{1}{2\sqrt{\pi c}}\int\limits_{-\infty}^{+\infty}df\exp\left[-\frac{f^2}{4c}+i\epsilon_{ij}\Sigma_{ij}
\left(f-\frac{g}{2}\left<E\right>\right)\right],
\end{equation}
where 

\begin{equation}
\label{c}
c\equiv\frac{g^2}{8}\left(\left<E^2\right>-\left<E\right>^2\right).
\end{equation} 
The last formula in~(\ref{A}) means that, instead 
of the constant field
$\left<E\right>$, we are now dealing with a shifted (but still space-independent) 
one $\bar E\equiv\left<E\right>-\frac{2f}{g}$,
where the field $f$ should be eventually averaged over. The respective pair-production rate reads

\begin{equation}
\label{neww}
w=\frac{N_cN_f}{2\sqrt{\pi c}}
\int\limits_{-\infty}^{+\infty}df{\rm e}^{-\frac{f^2}{4c}}\cdot
\frac{(g\bar E)^2}{(2\pi)^3}\sum\limits_{k=1}^{\infty}\frac{(-1)^{k+1}}{k^2}
\exp\left(-\frac{\pi k m^2}{g\bar E}\right).
\end{equation}
The parameter of the cumulant expansion, i.e. the ratio of the absolute value of the 
second cumulant to that of the first one, is

\begin{equation}
\label{par}
\frac{g}{2}\cdot 2\pi R_p^2\cdot \frac{\left<E^2\right>-\left<E\right>^2}{\left<E\right>}=g\left<E\right>
\cdot \pi R_p^2d.
\end{equation}
Here, 

\begin{equation}
\label{d}
d\equiv \frac{\left<E^2\right>-\left<E\right>^2}{\left<E\right>^2}
\end{equation} 
is a measure of dispersion of the field $E(r)$. It is calculated in Appendix~A and 
reads $d=\frac{\pi}{2}-1$. The condition of convergence of the cumulant expansion, i.e. the demand 
that the parameter~(\ref{par}) is smaller than unity, then yields yet another upper boundary on $L$: 

\begin{equation}
\label{Bound}
L<\frac{4}{\pi d}\frac{\sigma}{m^2}\simeq 2.23\frac{\sigma}{m^2},
\end{equation}
that is, however, weaker than 
the condition~(\ref{L}). As well as eq.~(\ref{L}), this new condition on $L$ becomes softer at $T\to T_c$, since 
$\frac{\sigma}{m^2}=\frac{\sigma_0}{m_\pi^2}t^{-1.11}\to\infty$. In another words, at a fixed $L$, 
obeying condition~(\ref{Bound}), cumulant expansion 
converges better at $T\to T_c$, since its parameter vanishes in this limit as ${\cal O}(t^{1.11})$.

To perform the average over $f$ in eq.~(\ref{neww}), note that the dispersion (i.e. the width) of 
the Gaussian distribution, ${\rm e}^{-ax^2}$, is $\frac{1}{\sqrt{2a}}$. Therefore, characteristic $f$'s
obey the estimate $|f|\le\sqrt{2c}=\frac{g}{2}\sqrt{\left<E^2\right>-\left<E\right>^2}$, and 
$\frac{2|f|/g}{\left<E\right>}\le\sqrt{d}\simeq\sqrt{0.57}
\simeq 0.76$. Therefore, although with the accuracy of only 76\% (cf. Section~5, where this problem will be avoided),
we can approximately write 

\begin{equation}
\label{barE}
\frac{1}{\bar E}\simeq\frac{1}{\left<E\right>}\left(1+\frac{2f}{g\left<E\right>}\right).
\end{equation}
When this expansion is substituted into eq.~(\ref{neww}), the $f$-integration can already be performed and yields
[cf. eq.~(\ref{w})]

\begin{equation}
\label{Wa}
w\simeq N_cN_f\frac{(g\left<E\right>)^2}{(2\pi)^3}\sum\limits_{k=1}^{\infty}\frac{(-1)^{k+1}}{k^2}
\left[1+\frac{2\pi m^2k}{g\left<E\right>}d\right]
\exp\left[-\frac{\pi m^2k}{g\left<E\right>}\left(1-\frac{\pi m^2k}{2g\left<E\right>}d\right)\right].
\end{equation} 
Approximating again the whole sum by the first term only, we see that the obtained correction is small, provided 
$\frac{\pi m^2}{g\left<E\right>}d\ll 1$.
This is precisely the condition of convergence of the cumulant expansion, eq.~(\ref{Bound}), with ``$<$'' 
replaced by ``$\ll$''. In particular, 
the correction produced by the second cumulant becomes vanishingly smaller than 
the leading term at $T\to T_c$. We also see that 
the obtained correction increases $w$, thus diminishing $\bar r$. Due to condition~(\ref{L}), 
$\bar r$ diminishes only by a factor of the order of unity.

\section{Accounting for quantum effects by the Feynman variational method}
In this Section, we will evaluate the leading quantum correction to eq.~(\ref{Wa}). It can be obtained upon  
the small-$T$ analysis of the path integral over ${\bf x}_\perp(\tau)$ in eq.~(\ref{1}) with the approximation~(\ref{smallT})
adopted. Such an integral can naturally be evaluated using the Feynman variational method~\cite{2}. 
In 2d-case and in our notations, it looks as follows.
We need to evaluate the path integral

$$
{\cal Z}(T)\equiv
\int d^2x(0)\int\limits_{{\bf x}(0)={\bf x}(T)}^{}{\cal D}{\bf x}(\tau){\rm e}^{-{\cal S}},$$
where ${\cal S}\equiv\int\limits_{0}^{T}d\tau
\left(\frac{\dot{\bf x}^2}{4}+U({\bf x})\right)$,
$U({\bf x})\equiv\frac{ig}{2T}E({\bf x})\epsilon_{ij}\Sigma_{ij}$. The 
classical expression for this integral, 
$\frac{1}{4\pi T}\int d^2x{\rm e}^{-TU({\bf x})}$, given by eq.~(\ref{class}), corresponds to $T$'s, which are 
so small that the trajectory does not deviate from its initial point ${\bf x}(0)$. Let us further 
introduce the coordinate describing the position of the trajectory ${\bf x}_0=\frac1T\int\limits_{0}^{T}d\tau{\bf x}(\tau)$,
the trial action ${\cal S}_0=\int\limits_{0}^{T}d\tau\frac{\dot{\bf x}^2}{4}+TW({\bf x}_0)$, and the respective trial partition function

$$
{\cal Z}_0(T)=\int d^2x_0\int\limits_{{\rm fixed}{\,}{\bf x}_0}^{}{\cal D}{\bf x}{\rm e}^{-{\cal S}_0}=$$

\begin{equation}
\label{Z0}
=\int d^2x_0\int\limits_{{\bf y}(0)={\bf y}(T)=0}^{}{\cal D}{\bf y}{\rm e}^{-\int\limits_{0}^{T}d\tau
\frac{\dot{\bf y}^2}{4}-TW({\bf x}_0)}=\frac{1}{4\pi T}\int d^2x{\rm e}^{-TW({\bf x})},
\end{equation}
where ${\bf y}(\tau)={\bf x}(\tau)-{\bf x}(0)$.
Here $W$ is a trial function, which 
should be determined upon the minimization of the expression $F_0+\frac1T\left<{\cal S}-{\cal S}_0\right>_{{\cal S}_0}$  (which approximates 
the true free energy $F$ of the system from the above, $F\le F_0+\frac1T\left<{\cal S}-{\cal S}_0\right>_{{\cal S}_0}$), where 
$F=-\frac1T\ln{\cal Z}(T)$, $F_0=-\frac1T\ln{\cal Z}_0(T)$. It is further possible to demonstrate that the
averaged difference of actions can be written as 

$$\frac1T\left<{\cal S}-{\cal S}_0\right>_{{\cal S}_0}=\left<U({\bf x}(0))\right>_{{\cal S}_0}-\left<W({\bf x}_0)\right>_{{\cal S}_0}.$$
For the first of the two averages here one has $\left<U({\bf x}(0))\right>_{{\cal S}_0}=\int d^2y{\cal K}({\bf y})
{\rm e}^{-TW({\bf y})}$, where 

$$
{\cal K}({\bf y})\equiv\int\limits_{{\bf x}(0)={\bf x}(T)}^{}{\cal D}{\bf x}U({\bf x}(0))\exp\left(-\int\limits_{0}^{T}d\tau
\frac{\dot{\bf x}^2}{4}\right)\delta({\bf y}-{\bf x}_0)=$$

$$
=\int\frac{d^2q}{(2\pi)^2}\tilde U({\bf q})\int\frac{d^2k}{(2\pi)^2}{\rm e}^{-i{\bf k}{\bf y}}
\int\limits_{{\bf x}(0)={\bf x}(T)}^{}{\cal D}{\bf x}\exp\left[-\int\limits_{0}^{T}d\tau
\frac{\dot{\bf x}^2}{4}+\frac{i}{T}{\bf k}\int\limits_{0}^{T}d\tau{\bf x}+i{\bf q}{\bf x}(0)\right],$$
and $\tilde U({\bf q})\equiv\int d^2x{\rm e}^{-i{\bf q}{\bf x}}U({\bf x})$ is the Fourier image of $U({\bf x})$.
Using the formula

$$\int\limits_{{\bf x}(0)={\bf x}(T)}^{}{\cal D}{\bf x}\exp\left[-\int\limits_{0}^{T}d\tau\left(
\frac{\dot{\bf x}^2}{4}+i{\bf f}{\bf x}\right)\right]=$$

$$=\frac{\pi}{T}\delta\left(\int\limits_{0}^{T}
d\tau{\bf f}\right)\exp\left[\frac12\int\limits_{0}^{T}d\tau\int\limits_{0}^{T}d\tau'|\tau-\tau'|
{\bf f}(\tau){\bf f}(\tau')+\frac{1}{T}\sum\limits_{\alpha=1}^{2}\left(\int\limits_{0}^{T}d\tau\tau f_\alpha
\right)^2\right]$$
with ${\bf f}=\frac{{\bf k}}{T}+{\bf q}\delta(\tau)$, one obtains 
${\cal K}({\bf y})=\frac{3}{(2\pi T)^2}\int d^2xU({\bf x}){\rm e}^{-\frac{3}{T}({\bf x}-{\bf y})^2}$. 
Thus, according to eq.~(\ref{Z0}),

$$\frac1T\left<{\cal S}-{\cal S}_0\right>_{{\cal S}_0}=\frac{\int d^2y{\rm e}^{-TW({\bf y})}\left[4\pi T{\cal K}({\bf y})-W({\bf y})
\right]}{\int d^2y{\rm e}^{-TW({\bf y})}}.$$
Note that the potential smeared over the Gaussian distribution,
$4\pi T{\cal K}({\bf y})$, goes over to the unsmeared potential $U({\bf y})$ in the limit $T\to 0$, as it should be.

The variational equation 
$\delta\left[F_0+\frac1T\left<{\cal S}-{\cal S}_0\right>_{{\cal S}_0}\right]=0$ then yields $W({\bf y})=4\pi T{\cal K}({\bf y})$
as the best choice of $W({\bf y})$~\footnote{With this choice of $W({\bf y})$, $\left<{\cal S}-{\cal S}_0\right>_{{\cal S}_0}=0$.}. 
Accordingly, the new expression, which accounts for quantum effects, 

\begin{equation}
\label{z0}
{\cal Z}_0(T)=\frac{1}{4\pi T}\int d^2y{\rm e}^{-4\pi T{\cal K}({\bf y})}=
\frac{1}{4\pi T}\int d^2y\exp\left[-\frac{3}{\pi}\int d^2xU({\bf x}){\rm e}^{-\frac{3}{T}({\bf x}-{\bf y})^2}\right]
\end{equation}
is a better approximation to ${\cal Z}(T)$ than its purely classical counterpart 
$\frac{1}{4\pi T}\int d^2x{\rm e}^{-TU({\bf x})}$
(recoverable in the limit $T\to 0$), which was used before.
 
The obtained result~(\ref{z0}) prescribes to replace $E({\bf x}_\perp)$ in eq.~(\ref{0}) by 

$${\cal E}({\bf x}_\perp)\equiv\frac{3}{\pi T}\int d^2yE({\bf y})
{\rm e}^{-\frac{3}{T}({\bf y}-{\bf x}_\perp)^2}.$$ 
In particular,
$\left<{\cal E}\right>=\left<E\right>$, i.e., at the level of the 
first cumulant, the variational method yields the same result as the clasical approximation.
We further obtain

$$\left<{\cal E}^2\right>=\frac{3}{2\pi TS}
\int d^2xd^2y{\rm e}^{-\frac{3}{2T}({\bf x}-{\bf y})^2}
E({\bf x})E({\bf y}).$$ 
This expression replaces $\left<E^2\right>$ in eq.~(\ref{A});
in particular, $\left<{\cal E}^2\right>\stackrel{T\to 0}{\longrightarrow}\left<E^2\right>$. 
Thus, within the variational approach, the constant~(\ref{c}) becomes replaced by
$c_{\rm var}\equiv\frac{g^2}{8}\left(\left<{\cal E}^2\right>-\left<E\right>^2\right)$.
An apparent difference of 
$c_{\rm var}$ from $c$ is that the former is $T$-dependent, whereas the latter is not~\footnote{In particular,
this means that the $f$-integral in eq.~(\ref{A}) should now be considered only inside the $T$-integral (and not vice versa).}.
The one-loop effective action reads 

$$\Gamma[A_i]\simeq S\frac{N_cN_f}{32\pi^{5/2}}\int\limits_{0}^{\infty}\frac{dT}{T^2}{\rm e}^{-m^2T}
\frac{1}{\sqrt{c_{\rm var}}}\int\limits_{-\infty}^{+\infty}df{\rm e}^{-\frac{f^2}{4c_{\rm var}}}
\frac{g\bar E}{\sin(g\bar E T)},$$
where ``$\simeq$'' means ``in the bilocal approximation to the cumulant expansion'',
and again $\bar E\equiv\left<E\right>-\frac{2f}{g}$.

Let us further calculate the integral entering $c_{\rm var}$

\begin{equation}
\label{intl}
\int d^2xd^2y{\rm e}^{-\frac{3}{2T}({\bf x}-{\bf y})^2}K_0(m_V|{\bf x}|)K_0(m_V|{\bf y}|)=
\frac{1}{m_V^4}\int d^2zd^2z'{\rm e}^{-\frac{3}{2Tm_V^2}({\bf z}-{\bf z}')^2}K_0(z)K_0(z'),
\end{equation}
where $z\equiv |{\bf z}|$, $z'\equiv |{\bf z}'|$.
This can be done approximately, by using the fact that $T<\frac{1}{m^2}\ll\frac{1}{m_V^2}$, i.e. 
$\frac{1}{Tm_V^2}\gg 1$, that allows us to Taylor expand $K_0(z')$ around ${\bf z}$. This calculation, whose details are given in Appendix~A,
leads to the following result:

\begin{equation}
\label{cvar}
c_{\rm var}=c+\frac{\sigma^3}{6g^2L^3}\frac{y}{g\bar E},
\end{equation}
where the variable $y\equiv g\bar ET$ acquires the values $\pi k$ when one is taking ${\rm Im}{\,}\Gamma[A_i]$.
We can further use the approximation~(\ref{barE}) with the term $\frac{2f}{g\left<E\right>}$ neglected, since 
it would otherwise lead to the excess of accuracy, making the $f$-integral non-Gaussian. Denoting this approximation by
``$\simeq$'' and substituting the above-mentioned values of $y$, we obtain the following modification of eq.~(\ref{c}):

$$c_{\rm var}\simeq\frac{g^2}{8}\left[\left(1+\frac{\pi k}{48g^2}\right)\left<E^2\right>-\left<E\right>^2\right].$$
This finally results in the following change of the dispersion parameter (\ref{d}), entering eq.~(\ref{Wa}),
that makes this parameter $k$-dependent:

$$d\to d_k\equiv\frac{\left(1+\frac{\pi k}{48g^2}\right)\left<E^2\right>-\left<E\right>^2}{\left<E\right>^2}.$$
This formula describes the quantum correction to eq.~(\ref{Wa}). We, however, see that this correction is  
small. Indeed, due to the main exponential factor in eq.~(\ref{Wa}), we have for relevant $k$'s:

$$
\frac{\pi k}{48g^2}<\frac{\pi}{48g^2}\cdot \frac{g\left<E\right>}{\pi m^2}=\frac{\sigma}{12L(gm)^2}\ll\frac{1}{48\pi},$$
where the last inequality stems from the condition $m\gg m_V$ [expressed by the left inequality of~(\ref{ineq})].

\section{Accounting for the space dependence of $E(r)$ without the cumulant expansion}
In this Section, we will evaluate $w$ in an alternative way, namely completely
without the use of cumulant expansion in eq.~(\ref{0}). The necessity of this calculation 
is, firstly, because cumulant expansion has been shown to converge only provided $L$ is bounded from above 
according to the condition~(\ref{Bound}). The new method enables one to replace this constraint 
by some other, weaker, one, and to relax the constraint~(\ref{L}). Secondly, the approximation used 
to arrive at eq.~(\ref{Wa}), which was holding with only 76\% accuracy, will now be not necessary anymore.

The method of this Section is based on 
averaging every term in the expansion of the exponent in eq.~(\ref{0})
over $d^2x_\perp$. In the London limit, such an average can be done analytically by making use of an explicit form of $E(r)$. 
Namely, we have

$$
\int
d^2x_\perp\exp\left[-\frac{ig}{2}\epsilon_{ij}\Sigma_{ij}E({\bf x}_\perp)\right]=
\sum\limits_{n=0}^{\infty}\frac{1}{n!}\left(-\frac{ig}{2}\epsilon_{ij}\Sigma_{ij}\right)^n
\left(\frac{m_V^2}{g_m}\right)^n\int d^2x_\perp(K_0(m_Vr))^n.$$
The last integral approximately equals $\frac{\pi}{m_V^2}2^{2-n}n!$, that
yields

$$
\frac{4\pi}{m_V^2}\sum\limits_{n=0}^{\infty}\left(-\frac{ig}{4}\frac{m_V^2}{g_m}
\epsilon_{ij}\Sigma_{ij}\right)^n=\frac{4\pi/m_V^2}{1+\frac{ig^2m_V^2}{8\pi}\epsilon_{ij}\Sigma_{ij}}
=\frac{4\pi}{m_V^2}\int\limits_{0}^{\infty}dt{\rm
e}^{-t\left(1+\frac{ig^2m_V^2}{8\pi}\epsilon_{ij}\Sigma_{ij}\right)}.
$$
The series above converges at $\frac{g^2m_V^2}{8\pi}\cdot 2\pi R_p^2<1$, that
results in the following boundary on $L$ from above [cf. eq.~(\ref{Bound})]: 

\begin{equation}
\label{LLL}
L<\frac{16}{\pi}\frac{\sigma}{m^2}\simeq 5.09\frac{\sigma}{m^2}.
\end{equation}
We further obtain for eq.~(\ref{eff}):

$$\Gamma[A_i]\simeq S\frac{N_cN_f}{\pi}\int\limits_{0}^{\infty}\frac{dT}{T^2}{\rm e}^{-m^2T}\int {\cal D}{\bf x}_\parallel
\int\limits_{0}^{\infty}dt{\rm
e}^{-t\left(1+\frac{ig^2m_V^2}{8\pi}\epsilon_{ij}\Sigma_{ij}\right)}=$$

\begin{equation}
\label{newG}
=S\frac{N_cN_f}{4\pi^2}\int\limits_{0}^{\infty}\frac{dT}{T^2}{\rm e}^{-m^2T}
\int\limits_{0}^{\infty}dt{\rm e}^{-t}\frac{g\bar E}{\sin(g\bar E T)},
\end{equation}
where $\bar E\equiv\frac{tgm_V^2}{4\pi}$ is a new space-independent electric field. We see that
the constraint~(\ref{L}) is now removed. Indeed, the condition for the first pole in 
${\rm Im}{\,}\Gamma[A_i]$ to contribute, $\pi\le g\bar ET$, together with the inequality $T<\frac{1}{m^2}$
yield $\pi<\frac{g\bar E}{m^2}=\frac{t}{L}\frac{\sigma}{m^2}$. 
This condition does not produce anymore a constraint on $L$, but 
merely means that $t$'s obeying the inequality 
$t>\pi L\frac{m^2}{\sigma}$ give a dominant contribution to ${\rm Im}{\,}\Gamma[A_i]$.

The pair-creation rate stemming from eq.~(\ref{newG}) takes the form

$$
w=\frac{N_cN_f}{2\pi^3}\left(\frac{\sigma}{L}\right)^2\sum\limits_{k=1}^{\infty}
\frac{(-1)^{k+1}}{k^2}\int\limits_{0}^{\infty}dtt^2{\rm e}^{-t-\frac{\pi
Lm^2k}{\sigma t}}=$$

$$=N_cN_f\frac{m^3}{\pi^{3/2}}\sqrt{\frac{\sigma}{L}}\sum\limits_{k=1}^{\infty}
\frac{(-1)^{k+1}}{\sqrt{k}}
K_3\left(2m\sqrt{
\frac{\pi Lk}{\sigma}}\right).$$
Clearly, only terms with $k\le\frac{\sigma}{4\pi Lm^2}$ are relevant in the
sum. At $T=0$,
substituting $\sigma=(440{\,}{\rm MeV})^2$ and a typical value $m=200{\,}{\rm MeV}$,
we obtain $k<\frac1L<1$. Therefore, 
only the first term is relevant, that yields

\begin{equation}
\label{modw}
w\simeq N_cN_f\frac{m^3}{\pi^{3/2}}\sqrt{\frac{\sigma}{L}}
K_3\left(2m\sqrt{
\frac{\pi L}{\sigma}}\right).
\end{equation}
According to the above-obtained constraint~(\ref{LLL}), the
argument of the MacDonald function in this formula is smaller than 8. For the values of $L$, at which this
argument is still larger than unity, i.e. $L>\frac{\sigma}{4\pi m^2}$,

\begin{equation}
\label{Neww}
w\simeq N_cN_f\frac{m^{5/2}\sigma^{3/4}}{2\pi^{5/4}L^{3/4}}{\rm
e}^{-2m\sqrt{\frac{\pi L}{\sigma}}}.
\end{equation}
When $L$ obeys simultaneously inequality~(\ref{L}), the obtained expression can be compared to eq.~(\ref{ww}).
The parametric dependences of these two expression on $m$, $\sigma$, and $L$ are
apparently different from each other. The regime under
discussion, $2m\sqrt{\frac{\pi L}{\sigma}}> 1$, however, does not imply the exponential smallness of~(\ref{Neww}),
since, due to~(\ref{L}), $2m\sqrt{\frac{\pi L}{\sigma}}<4$. The string-breaking distance, corresponding to eq.~(\ref{Neww}),
reads 

\begin{equation}
\label{newdist}
\bar r=\frac{2\pi^{5/8}}{g\sqrt{N_cN_f}}\left(\frac{\sigma}{L}\right)^{1/8}\frac{1}{m^{5/4}}
{\rm e}^{m\sqrt{\frac{\pi L}{\sigma}}}.
\end{equation}
Its ratio to distance~(\ref{rr}) is given by the function 
$\frac{2}{\pi^{1/4}x^{5/4}}{\rm e}^{x-\frac{x^2}{8}}$, where the variable $x\equiv m\sqrt{\frac{\pi L}{\sigma}}$
ranges between $\frac12$ and 2. This is a monotonically decreasing function, which therefore acquires its maximum 
at $x=\frac12$ [where approximation~(\ref{Neww}) to eq.~(\ref{modw}) starts breaking down].
The value of the maximum, $\simeq 9.10$, 
is, thus, an upper limit for the ratio of the two string-breaking distances. The minimal value of this  
ratio, corresponding to $x=2$, is only $\simeq 2.84$.

Instead, at $T\to T_c$,
$\frac{m}{\sqrt{\sigma}}=\frac{m_\pi}{\sqrt{\sigma_0}}t^{0.55}\to 0$, and 
we have $w\to\frac{N_cN_f}{\pi^3}\left(\frac{\sigma}{L}\right)^2$, that differs from eq.~(\ref{tto0}) only by a factor 2.
The respective string-breaking distance is, therefore, larger only by a factor $\sqrt{2}$.

\section{Considering pairs as holes in the confining pellicle}
In this Section, we will consider an alternative approach to the pair production, based on a combination of 
the formulae on the metastable vacuum decay~\cite{5,6} with the stochastic vacuum model~\cite{7} (for a recent review 
see~\cite{8}). The idea is to consider the produced pairs as holes in a 2d confining pellicle, which spans over the 
contour of an external $q\bar q$-pair. Such a hole is a region where the pellicle is eaten up. 
Therefore, a hole of a radius $R$ diminishes the action of the pellicle by $\sigma\cdot\pi R^2$,
but increases this action by $(m+\mu)\cdot 2\pi R$. 
Here $\mu$ is some parameter of dimension [mass], which is
a nonperturbative part of a constant entering the 
perimeter law of the small-sized Wilson loop of a produced pair.
The critical radius (i.e. such a radius, that all holes with $R<R_c$ collapse, while those
with $R>R_c$ expand and destroy the pellicle) stems therefore from extremization of the action 
$S(R)\equiv(m+\mu)\cdot 2\pi R-\sigma\cdot\pi R^2$. This
critical radius and the action of a critical hole are $R_c=(m+\mu)/\sigma$ and $S(R_c)=\pi(m+\mu)^2/\sigma$. 
Accordingly, in this approach, 
the rate of the pair production is proportional to ${\rm e}^{-\pi(m+\mu)^2/\sigma}$. The proportionality 
coefficient~\cite{6}, 
$\frac{\sigma}{2\pi}$, should furthermore be multiplied by the factor $N_f$~\footnote{The factor $N_c$ is now absent, since 
the Wilson loop, describing a produced pair, is a colorless object. Instead, the number of Wilson loops, which 
can potentially be created, is proportional to the number of different quark species, i.e. to $N_f$.}:

\begin{equation}
\label{newW}
w=N_f\frac{\sigma}{2\pi}\exp\left[-\frac{\pi m^2}{\sigma}\left(
1+\frac{\mu}{m}\right)^2\right].
\end{equation} 
Next, to evaluate both $\mu$ and $\sigma$ within the 
same model, it is natural to use the stochastic vacuum model. It yields the following expression for the Wilson loop
$\left<W(C)\right>_{\rm YM}\equiv\frac{1}{N_c}
\left<{\rm tr}{\,}{\cal P}\exp\left(ig\oint\limits_{C}^{}
dx_\mu A_\mu\right)\right>_{\rm YM}$:

\begin{equation}
\label{Wloop}
\left<W(C)\right>_{\rm YM}\simeq
\frac{1}{N_c}{\,}{\rm tr}{\,}\exp\left[-\frac{1}{2!}\frac{g^2}{4}\int\limits_{\Sigma(C)}^{}d\sigma_{\mu\nu}(x)
\int\limits_{\Sigma(C)}^{}d\sigma_{\lambda\rho}(x')\left<F_{\mu\nu}(x)\Phi_{xx'}F_{\lambda\rho}(x')\Phi_{x'x}\right>_{\rm YM}
\right],
\end{equation}
where $\Sigma(C)$ is the surface encircled by the flat contour $C$. Next, $A_\mu\equiv A_\mu^aT^a$, where $T^a$'s
stand for the generators of the group SU($N_c$) in the fundamental representation,
$\left[T^a,T^b\right]=if^{abc}T^c$, ${\rm tr}~ T^aT^b=\frac12\delta^{ab}$; the average $\left<\ldots\right>_{\rm YM}$
is implied with respect to the Euclidean Yang-Mills action, $\frac14\int d^4x(F_{\mu\nu}^a)^2$, where 
$F_{\mu\nu}^a=\partial_\mu A_\nu^a-\partial_\nu A_\mu^a+gf^{abc}A_\mu^b A_\nu^c$, $a=1, \ldots, N_c^2-1$;
$\Phi_{xx'}\equiv\frac{1}{N_c}{\,}{\cal P}{\,}\exp\left(ig\int\limits_{x'}^{x}dz_\mu A_\mu(z)\right)$ is a phase 
factor along the straight line, which goes through $x'$ and $x$. 
The symbol ``$\simeq$'' in eq.~(\ref{Wloop})
is implied in the sense of the bilocal approximation to the cumulant expansion. This approximation, supported 
by the lattice data~\cite{8, 9, 10}, states that,
in the Yang-Mills theory, the 
two-point irreducible gauge-invariant correlation function (cumulant) of $F_{\mu\nu}$'s dominates over all cumulants 
of higher orders, which are therefore neglected. Finally,
the factor $1/2!$ in eq.~(\ref{Wloop}) 
is simply due to the cumulant expansion, whereas the factor
$1/4$ is due to the non-Abelian Stokes' theorem.

The stochastic vacuum model suggests further the following parametrization of the two-point cumulant:

$$\left<F_{\mu\nu}(x)\Phi_{xx'}F_{\lambda\rho}(x')\Phi_{x'x}\right>_{\rm YM}=\frac{\hat 1_{N_c\times N_c}}{N_c}{\cal N}
\Biggl\{(\delta_{\mu\lambda}\delta_{\nu\rho}-\delta_{\mu\rho}\delta_{\nu\lambda})D\left((x-x')^2\right)+$$

$$
+\frac12\left[\partial_\mu^x\left((x-x')_\lambda\delta_{\nu\rho}-(x-x')_\rho\delta_{\nu\lambda}\right)+
\partial_\nu^x\left((x-x')_\rho\delta_{\mu\lambda}-(x-x')_\lambda\delta_{\mu\rho}\right)\right]\times$$

\begin{equation}
\label{pa}
\times\left[D_1\left((x-x')^2\right)+\frac{N_cC_2}{{\cal N}\pi^2|x-x'|^4}\right]
\Biggr\}.
\end{equation}
Here $C_2\equiv\frac{N_c^2-1}{2N_c}$ is the quadratic Casimir operator of the fundamental representation, and 
${\cal N}$ is the normalization constant, which in 4d reads 
${\cal N}=\frac{\left<(F_{\mu\nu}^a)^2\right>_{\rm YM}}{24[D(0)+D_1(0)]}$~\footnote{Note that the one-gluon-exchange 
contribution, represented by the $\frac{1}{|x-x'|^4}$-term, can alternatively be considered as a part of the 
function $D_1$~\cite{8}. Here, we rather consider this perturbative contribution separately, so that $D_1(0)$ is a finite
quantity.}. Inserting this parametrization into
eq.~(\ref{Wloop}) we obtain (applying Abelian Stokes' theorem to the part containing derivatives):

$$\left<W(C)\right>_{\rm YM}\simeq\exp\Biggl\{-\frac{g^2{\cal N}}{8N_c}\Biggl[2
\int\limits_{\Sigma(C)}^{}d\sigma_{\mu\nu}(x)
\int\limits_{\Sigma(C)}^{}d\sigma_{\mu\nu}(x')D\left((x-x')^2\right)+$$

\begin{equation}
\label{G}
+\oint\limits_{C}^{}dx_\mu
\oint\limits_{C}^{}dx'_\mu\left[G\left((x-x')^2\right)+\frac{N_cC_2}{{\cal N}\pi^2}\frac{1}{(x-x')^2}\right]
\Biggr]\Biggr\},
\end{equation}
where $G(x^2)\equiv\int\limits_{x^2}^{\infty}dt D_1(t)$. The one-gluon-exchange contribution to the Wilson loop,
$\exp\left[-\frac{g^2C_2}{8\pi^2}\oint\limits_{C}^{}dx_\mu
\oint\limits_{C}^{}dx'_\mu\frac{1}{(x-x')^2}\right]\simeq\exp\left(-\frac{g^2C_2}{8\pi a}\right)$ is known \cite{12} to yield the 
renormalization of mass of a produced pair. Here $a$ stands for an inverse UV cutoff; $a\ll T_g$, and $T_g\simeq 1{\,}{\rm GeV}^{-1}$
is the correlation length of the vacuum~\cite{7, 8, 9, 10}.
The asymptotics of the Wilson loop at $\sqrt{|\Sigma(C)|}\gg T_g$, where $|\Sigma(C)|$ is the area of $\Sigma(C)$,
is $\left<W(C)\right>_{\rm YM}\simeq{\rm e}^{-\sigma|\Sigma(C)|}$. This asymptotics is obeyed by the Wilson loop
of an external $q\bar q$-pair. Instead, at  $\sqrt{|\Sigma(C)|}={\cal O}(T_g)$, i.e. for the Wilson loop of a produced pair,
$\left<W(C)\right>_{\rm YM}\simeq
{\rm e}^{-(m+\mu) L(C)}$, where $L(C)$ is the length of $C$~\footnote{Below, we will see that the 
typical size of the Wilson loop of a produced pair is indeed ${\cal O}(T_g)$.}.
As follows from eq.~(\ref{G}), the parametrization~(\ref{pa}) 
is chosen in such a way that the function $D$ yields the area law,
while the term with the function $D_1$ contributes to the perimeter law. Namely, 
the string tension reads~\cite{11}
$\sigma=\frac{g^2{\cal N}}{2N_c}\int d^2x D({\bf x}^2)$,
while for the perimeter constant $\mu$ we obtain in a way similar to the Coulomb \cite{12} 
interaction between points lying on $C$: 
$\mu=\frac{g^2{\cal N}}{8N_c}\int\limits_{0}^{\infty}d\xi G(\xi^2)$. A derivation of this formula is presented in Appendix~A.

In what follows, we will adopt the exponential   
parametrization of the functions $D$ and $D_1$~\cite{8, 9, 10}
$D(x^2)=D(0){\rm e}^{-|x|/T_g}$, $D_1(x^2)=D_1(0){\rm e}^{-|x|/T_g}$. 
It yields the following values of the string tension and the perimeter constant:
$\sigma=\frac{\pi g^2{\cal N}}{N_c}T_g^2D(0)$,
$\mu=\frac{g^2{\cal N}}{2N_c}T_g^3D_1(0)$. Introducing the 
parameter $\gamma\equiv D_1(0)/D(0)$, whose lattice value in full QCD is $0.13\pm 0.08$ (see e.g.~\cite{14}),
we can rewrite the obtained results as

\begin{equation}
\label{sigma}
\sigma=\frac{\pi}{24(1+\gamma)N_c}g^2\left<(F_{\mu\nu}^a)^2\right>T_g^2,~~ 
\mu=\frac{\gamma}{48(1+\gamma)N_c}g^2\left<(F_{\mu\nu}^a)^2\right>T_g^3.
\end{equation}
For the critical radius of a hole we have $R_c=\frac{m+\mu}{\sigma}=\frac{m}{\sigma}+
\frac{\gamma}{2\pi}T_g$. Substituting again $\sigma=(440{\,}{\rm MeV})^2$, $m=200{\,}{\rm MeV}$, and 
$T_g^{-1}=1{\,}{\rm GeV}$, 
we get for the worst case of the 
above-quoted lattice value of $\gamma$, $\gamma=0.21$, $\frac{R_c}{T_g}\simeq 1.07$~\footnote{Notice that
$T_g$ itself is smaller (by a factor of the order of 5) 
than a typical size of the Wilson loop of an external $q\bar q$-pair, at which the 
onset of string-breaking is normally expected.}. 
This ratio is certainly ${\cal O}(1)$,
that justifies the use of the perimeter asymptotics for the Wilson loop of a produced pair. 

Comparing the pair production rate, eq.~(\ref{newW}), with eq.~(\ref{ww}), we see that the parameter 
$\left(1+\frac{\mu}{m}\right)^2$ replaces the parameter $L/4$ of that equation. At 
$\sigma=(440{\,}{\rm MeV})^2$, $m=200{\,}{\rm MeV}$, we have~\footnote{The correction $\frac{\mu}{m}$ is very small:
$\frac{\mu}{m}=\frac{\gamma}{2\pi}\frac{T_g\sigma}{m}\simeq 0.02$.}
$\left(1+\frac{\mu}{m}\right)^2\simeq 1.08$, whereas,
according to eq.~(\ref{L}), $\frac{L}{4}<1.55$, that is of the same order of magnitude.
Finally, the new string-breaking distance, at which the potential 
$V(r)=\sigma r{\rm e}^{-\pi r^2\cdot w}$ acquires its maximum, is 

\begin{equation}
\label{NEWr}
\bar r=\frac{1}{\sqrt{2\pi w}}=\frac{1}{\sqrt{N_f\sigma}}\exp\left[\frac{\pi m^2}{\sigma}\left(
1+\frac{\mu}{m}\right)^2\right].
\end{equation}
The argument of the exponent here approximately equals $0.70$, that is quite similar to $0.5$ 
we had as an upper boundary in case of eq.~(\ref{rr}).
However, contrary to that equation, eq.~(\ref{NEWr}) does not contain the factor $\sqrt{L}$ in the preexponent.  

Notice also that, 
due to the fact that $\sigma={\cal O}(N_c^0)$, the string-breaking distance~(\ref{rr}) is ${\cal O}(N_c^0)$ too~\cite{1}.
For the same reason, as can be seen from the first of eqs.~(\ref{sigma}),
$T_g$ is ${\cal O}(N_c^0)$ as well. 
Therefore, according to the second equation of~(\ref{sigma}), $\mu$ is also ${\cal O}(N_c^0)$.
This means that, as well as eq.~(\ref{rr}), the new string-breaking distance~(\ref{NEWr}) is ${\cal O}(N_c^0)$.

\section{Summary}
In this paper, we have considered two approaches to the problem of string breaking in QCD: one of these is based 
on the dual superconductor model of confinement, and the other one -- on the stochastic vacuum model.
In the first approach, which was proposed already in ref.~\cite{1}, the pair-production mechanism is due to the 
field of the chromoelectric flux tube (dual Abrikosov-Nielsen-Olesen string in the London limit of the 
dual Abelian Higgs model). In the second approach,
pairs are considered as holes in a pellicle, which confines a test $q\bar q$-pair. 

Within the first approach,
we have demonstrated that the result of ref.~\cite{1}, based on the Schwinger formula, accounts only
for the first term of the cumulant expansion in the average~(\ref{A}). In Section~3, we have found the temperature
dependences of the string-breaking distance at temperatures close to the critical one and at low temperatures,
smaller than ${\cal O}(f_\pi)$. In both cases, the string-breaking distance increases with the increase of the 
temperature. We have also noticed that, in the case of spin-$\frac12$ quarks, antiperiodic boundary conditions for fermions
produce only corrections which fall off as the tail of the Gaussian distribution,
as long as $T<m_\pi/\sqrt{2}$. As a by-product, we have found in Appendix~B the temperature dependence of the string tension, which
reproduces correctly the zero-temperature value.
In Section~4, we have calculated (for scalar quarks) a
correction, generated by the second cumulant in the expansion~(\ref{A}), 
i.e. by the dispersion of the chromoelectric field in the direction 
transverse to the string. This effect slightly 
diminishes the string-breaking distance. Using the Feynman variational method, 
we have then derived in Section~4 the leading quantum 
correction to this effect, produced by the deviation of the trajectory of a pair from the classical one.
This effect further diminishes the string-breaking distance.

The effects of dispersion of the chromoelectric field are small
as long as the cumulant expansion is convergent, that is 
the case when the logarithm of the Landau-Ginzburg parameter is bounded from above 
according to~(\ref{Bound}). This constraint is, however, always obeyed as long as at least the first pole in the 
Schwinger formula gives its contribution, that leads to an even more severe constraint~(\ref{L}). Although both 
constraints have been shown to relax at $T\to T_c$, they do exist at $T=0$. This fact necessitates to perform the 
average~(\ref{A}) without the use of the cumulant expansion at all, that has been done in Section~6. As a result, the 
upper boundary on the logarithm of the Landau-Ginzburg parameter increases by a factor 4 [cf. eq.~(\ref{LLL})].
Furthermore, a novel formula~(\ref{modw}) for the rate of the pair production has been derived. At $T=0$,
some range of the Landau-Ginzburg parameter has been found, in which the string-breaking distance is 
larger than the one we had with the use of
the bilocal approximation to the cumulant expansion in a factor varying
between $2.84$ and $9.10$~\footnote{This result indicates that, at least at these values of the Landau-Ginzburg parameter, 
the effect produced by cumulants higher than the quadratic one is opposite and stronger than the result produced by 
the quadratic cumulant alone.}.
Instead, at $T\to T_c$, this factor is smaller, namely it equals $\sqrt{2}$. 
Notice that such an analytic average of the exponent~(\ref{A}) without the
use of the cumulant expansion was only possible due to the explicit form of the Abrikosov-Nielsen-Olesen solution in the 
London limit. Apparently, studies away from this limit will require a numerical analysis.  

Within the approach, which treats pairs as holes in the confining pellicle, the new quantity on which 
the string-breaking distance is dependent is the constant entering the perimeter law of the Wilson loop
of a produced pair.
For typical values of the hadronic mass, string tension, and the vacuum correlation length, this dependence is,
however, very weak. As for the dependence of the new expression for the string-breaking distance 
on the mass of a produced pair and on the string tension, it is the same as in the above-discussed 
case of the approach based on the 
Schwinger formula, where only the second cumulant is taken into account. Finally, the results 
for the string-breaking distance, obtained within all the above-mentioned approaches, are ${\cal O}(N_c^0)$ 
in the large-$N_c$ limit.

\section*{Acknowledgments}
One of us (D.A.) acknowledges the Alexander~von~Humboldt foundation for the financial support.
He would also like to thank the staffs of the Physics Departments of the University of Pisa and of the 
Humboldt University of Berlin for the cordial hospitality.  

\appendix
\section{Some technical details}
Let us first present evaluation of the dispersion parameter~(\ref{d}). Integrating both sides of the equality
$\frac{1}{2\pi}K_0(mr)=\int\frac{d^2p}{(2\pi)^2}\frac{{\rm e}^{i{\bf p}{\bf x}_\perp}}{p^2+m^2}$ over $d^2r$
we trivially have $\int d^2rK_0(mr)=\int\frac{d^2p}{p^2+m^2}\delta({\bf p})=\frac{1}{m^2}$. In the same way

\begin{equation}
\label{a1}
\int d^2rK_0^2(mr)=\int d^2r\int\frac{d^2pd^2q}{(2\pi)^2}
\frac{{\rm e}^{i({\bf p}+{\bf q}){\bf x}_\perp}}{(p^2+m^2)(q^2+m^2)}=\frac{1}{2\pi}\int\frac{d^2p}{(p^2+m^2)^2}=
\frac{1}{2m^2}.
\end{equation}
The parameter~(\ref{d}) then reads

$$
d=\frac{\int d^2rK_0^2(m_Vr)-\frac{m_V^2}{\pi}\left[\int d^2rK_0(m_Vr)
\right]^2}{\frac{m_V^2}{\pi}\left[\int d^2rK_0(m_Vr)\right]^2}=\frac{\frac{1}{2m_V^2}-
\frac{1}{\pi m_V^2}}{\frac{1}{\pi m_V^2}}=\frac{\pi}{2}-1.$$

Next, we will discuss some details of derivation of $c_{\rm var}$. Taylor expanding $K_0(z')$
in eq.~(\ref{intl}) to the second order we have 

$$\frac{1}{m_V^4}\int d^2zd^2z'{\rm e}^{-\frac{3}{2Tm_V^2}({\bf z}-{\bf z}')^2}K_0(z)K_0(z')\simeq$$

$$
\simeq\frac{2\pi T}{3m_V^2}\int d^2z K_0^2(z)-\frac{1}{m_V^4}\int d^2z\frac{z_\mu}{z}K_1(z)
\int d^2z'(z'-z)_\mu{\rm e}^{-\frac{3}{2Tm_V^2}({\bf z}-{\bf z}')^2}+$$

$$+\frac{1}{2m_V^4}\int d^2z\left[
\frac{z_\mu z_\nu-z^2\delta_{\mu\nu}}{z^3}K_1(z)+\frac{z_\mu z_\nu}{2z^2}(K_0(z)+
K_2(z))\right]K_0(z)\times$$

$$\times\int d^2z'(z'-z)_\mu(z'-z)_\nu{\rm e}^{-\frac{3}{2Tm_V^2}({\bf z}-{\bf z}')^2}.$$
The second term on the r.h.s. of this equation apparently vanishes, while the third ones reads
$\frac{\pi T^2}{9}\int d^2zK_0\left[\frac12(K_0+K_2)-\frac{1}{z}K_1\right]$. Using the definition 
$c_{\rm var}\equiv\frac{g^2}{8}\left(\left<{\cal E}^2\right>-\left<E\right>^2\right)$, we then arrive at
the following intermediate result:

$$
c_{\rm var}=c+\frac{\pi\sigma^3T}{3g^2L^3}\int\limits_{0}^{\infty}dz K_0\left[
z(K_0+K_2)-2K_1\right].$$
Here, the addendum $c$, eq.~(\ref{c}), is apparently produced by the term with no derivatives, while the other addendum is produced 
by the second-derivative term of the Taylor expansion. Finally, the integral over $z$ can be calculated 
exactly. It reads

$$\frac{1}{2\pi}\int d^2z K_0^2+\int\limits_{0}^{\infty}dzK_0\left[z\left(K_0+\frac2z K_1\right)-2K_1\right]=
\frac{1}{\pi}\int d^2z K_0^2=\frac{1}{2\pi},$$
where eq.~(\ref{a1}) on the last step has been used. This yields eq.~(\ref{cvar}).

We will finally present a proof of the formula 

\begin{equation}
\label{length}
\oint\limits_{C}^{}dx_\mu\oint\limits_{C}^{}dx_\mu'
G\left((x-x')^2\right)\simeq L\cdot\int\limits_{0}^{\infty}d\xi G(\xi^2),
\end{equation} 
where $L=\int\limits_{0}^{1}ds
|\dot x_\mu(s)|$ is the length of the contour $C$. Since the function $G(x^2)$ is rapidly decreasing
[e.g. $G(x^2)=2D_1(0)T_g(|x|+T_g){\rm e}^{-|x|/T_g}$ for the adopted Ansatz $D_1(x^2)=D_1(0){\rm e}^{-|x|/T_g}$],
we can Taylor expand $x_\mu'\equiv x_\mu(s+t)$ as $x_\mu(s+t)\simeq x_\mu(s)+t\dot x_\mu(s)$. In the expression 
under study,
$\int\limits_{0}^{1}ds \int\limits_{0}^{1}dt\dot x_\mu(s)\dot x_\mu(s+t) G(t^2\dot x_\mu^2(s))$,
we therefore have $\dot x_\mu(s)\dot x_\mu(s+t)\simeq \dot x_\mu^2(s)+t\dot x_\mu(s)\ddot x_\mu(s)=
\dot x_\mu^2(s)$, where, at the last step, the proper-length parametrization $\dot x_\mu^2={\rm const}$ has been fixed.
Introducing instead of $t$ the new integration variable $\xi=t|\dot x_\mu|$, we have 
for the expression under study 
$\int\limits_{0}^{1}ds|\dot x_\mu(s)|\int\limits_{0}^{|\dot x_\mu|}d\xi G(\xi^2)$.
Finally, due to the rapid decrease of $G(\xi^2)$, the upper integration limit 
in the last integral can be replaced by infinity, that completes the proof of eq.~(\ref{length}).

\section{Temperature dependence of the string tension}
To find the dependence $\sigma(T)$, notice that the string tension can be derived from the 
string representation of the  
partition function~\cite{strrepr}. An essential result of 
this representation is the following string effective action:

$$2(\pi v)^2\int d\sigma_{\mu\nu}(x)\int d\sigma_{\mu\nu}(x')D_{m_V}(x-x'),$$ 
where $D_m(x)\equiv mK_1(m|x|)/(4\pi^2|x|)$
is the Yukawa propagator. The zero-temperature string tension then stems from this action according to the general (for this type
of non-local string actions) formula~\cite{11}: $\sigma_0=(\pi v)^2\cdot\frac{4}{m_V^2}
\int\limits_{m_V/m_H}^{}d^2z\frac{m_V^2K_1(|z|)}{4\pi^2|z|}$. This indeed equals the above-used value $2\pi v^2L$
(following from the Landau-Ginzburg equations). At finite temperatures (smaller than the temperature of dimensional reduction), 
we rather have the following 
expression in terms of Matsubara frequencies:

$$
\sigma(T)=v^2\int\limits_{m_V/m_H}^{}d^2z\sum\limits_{n=-\infty}^{+\infty}
\frac{K_1\left(\sqrt{z^2+(m_V\beta n)^2}\right)}{\sqrt{z^2+(m_V\beta n)^2}}.$$
To perform the integration, it is useful to transform 
the sum as follows:

$$\sum\limits_{n=-\infty}^{+\infty}
\frac{K_1\left(\sqrt{z^2+(m_V\beta n)^2}\right)}{\sqrt{z^2+(m_V\beta n)^2}}=\sum\limits_{n=-\infty}^{+\infty}
\int\limits_{0}^{\infty}dt\exp\left[-\frac{1}{4t}-t\left(z^2+(m_V\beta n)^2\right)\right]=$$

$$=\frac{T\sqrt{\pi}}{m_V}\sum\limits_{n=-\infty}^{+\infty}\int\limits_{0}^{\infty}\frac{dt}{\sqrt{t}}
\exp\left[-z^2t-\frac{1+(\omega_n/m_V)^2}{4t}\right]=
\frac{\pi T}{m_V|z|}\sum\limits_{n=-\infty}^{+\infty}{\rm e}^{-|z|\sqrt{1+(\omega_n/m_V)^2}}.$$
Integration of this expression over $d^2z$ is now trivial and yields 

$$
\sigma(T)=\frac{2\pi^2Tv^2}{m_V}\sum\limits_{n=-\infty}^{+\infty}\frac{{\rm e}^{-\frac{m_V}{m_H}\sqrt{1+(\omega_n/m_V)^2}}}
{\sqrt{1+(\omega_n/m_V)^2}}.$$
Performing now with the sum a transformation of the form~(\ref{transform}) in the opposite direction, we obtain

$$
\sigma(T)=2\pi v^2\sum\limits_{n=-\infty}^{+\infty}K_0\left(\frac{m_V}{m_H}\sqrt{1+(m_H\beta n)^2}\right).$$
In particular, at $T\to 0$, one may approximate the sum by the zeroth term, that recovers the value $\sigma_0=2\pi v^2 L$.


\end{document}